\documentclass[conference]{IEEEtran}
\usepackage{amsmath,amsfonts}
\usepackage{algorithm}
\usepackage{algpseudocode}
\usepackage{array}
\usepackage{subfigure}
\usepackage{textcomp}
\usepackage{stfloats}
\usepackage{url}
\usepackage{verbatim}
\usepackage{graphicx}
\usepackage{cite}
\usepackage[dvipsnames]{xcolor}
\usepackage{float}
\usepackage{hyperref}
\newcommand{\cmmnt}[1]{}
\begin{document}

\title{LLHR: Low Latency and High Reliability CNN Distributed Inference for Resource-Constrained UAV Swarms}
\author{
    \IEEEauthorblockN{Marwan Dhuheir\IEEEauthorrefmark{1},
    Aiman Erbad\IEEEauthorrefmark{1},
    Sinan Sabeeh\IEEEauthorrefmark{2}
    }
    \IEEEauthorblockA{\IEEEauthorrefmark{1}Division of Information and Computing Technology, College of Science and Engineering,\\ Hamad Bin Khalifa University, Qatar Foundation, Doha, Qatar.\\
    \IEEEauthorrefmark{2}Barzan Holdings QSTP LLC, Doha, Qatar.
   }
}
\maketitle

\begin{abstract}
Recently, Unmanned Aerial Vehicles (UAVs) have shown impressive performance in many critical applications, such as surveillance, search and rescue operations, environmental monitoring, etc. In many of these applications, the UAVs capture images as well as other sensory data and then send the data processing requests to remote servers. Nevertheless, this approach is not always practical in real-time-based applications due to unstable connections, limited bandwidth, limited energy, and strict end-to-end latency. One promising solution is to divide the inference requests into subtasks that can be distributed among UAVs in a swarm based on the available resources. Moreover, these tasks create intermediate results that need to be transmitted reliably as the swarm moves to cover the area. Our system model deals with real-time requests, aiming to find the optimal transmission power that guarantees higher reliability and low latency. We formulate the Low Latency and High-Reliability (LLHR) distributed inference as an optimization problem, and due to the complexity of the problem, we divide it into three subproblems. In the first subproblem, we find the optimal transmit power of the connected UAVs with guaranteed transmission reliability. The second subproblem aims to find the optimal positions of the UAVs in the grid, while the last subproblem finds the optimal placement of the CNN layers in the available UAVs. We conduct extensive simulations and compare our work to two baseline models demonstrating that our model outperforms the competing models.
\end{abstract}

\begin{IEEEkeywords}
Optimization; latency; UAVs positions; distributed inference; CNN network; classification decision; UAVs; reliability.
\end{IEEEkeywords}
\section{introduction}

Recently, Unmanned Aerial Vehicles (UAVs) have been widely used in various applications ranging from surveillance, search and rescue operations, last-mile delivery, provision of of wireless services, and precision agriculture \cite{bejiga2017convolutional,wu20205g,al2018survey}, leading to an increasing interest from the research community and the industry in UAV swarms. UAVs outperform traditional technologies owing to their high maneuverability, lower cost, easy control, and monitoring capabilities from low to level altitudes \cite{padro2019comparison}. 

All the data processing load for different requests can be handled locally within the swarm without resorting to remote servers. Nevertheless, the UAV swarm is susceptible to low reliability and high latency in the data transmission due to interference and the UAV mobility patterns during the mission. To address the reliability and latency challenges, Tengchan et al. in \cite{zeng2018wireless} have identified a latency threshold at which the reliability is assured; however, they did not consider the computational capacities of UAVs in their work. In this context, we derive the optimal power that guarantees reliable transmission between UAVs based on their computational capacities and the optimal positions of UAVs at which the interference and latency are at their minimum.

Recently, computer vision systems that analyze data captured by UAVs with deep neural networks (DNNs) have significantly improved \cite{al2018survey}. Furthermore, DNNs have become a state-of-the-art method for detecting and classifying objects in images \cite{yang2020offloading}. 
The DNNs contain millions of neurons and trillions of connections and require large amounts of computing resources to process inference tasks.
The distributed inference in multi-UAV systems is a promising way to partition the DNNs into a number of parts (segments or layers) and execute them to deal with the resource-constraints in UAVs. Each part of the model is assigned to one of the UAVs in the swarm. UAVs run their dedicated tasks and then share the output with the next participant until the final outcome is reached \cite{9498967}. 
Dealing with UAVs adds additional challenges as the data rates depends on the varying distances between devices. In previous works, the effect of mobility on the distributed inference in UAV swarms has not been considered, particularly while considering  exposure to path loss, interference, and possible disconnection.
Three different distributed inference scenarios exist for UAVs in surveillance-based applications. In the first approach, multiple UAVs capture images and send them to the CNN network in remote servers to execute it \cite{teerapittayanon2017distributed}. This scenario relies on the high capacity of remote machines that are not always adequate for online applications and are highly sensitive to delays; hence, it is inefficient in real-time-based applications where prompt intervention is required. The second approach divides the captured data into parts; some parts are sent to remote servers, while others are executed onboard. Yang et al. in \cite{yang2020offloading} is an example of this approach where the captured image is divided based on the quality of the image; higher quality parts are executed in remote servers while light parts are executed onboard. The third approach is to execute the inference across multiple UAVs in distributed inference to accomplish the surveillance mission \cite{9498967}. Their approaches aim to find the placement of the optimal layers within UAV participants to minimize the latency of the final classification decisions. However, these works do not consider the optimization of UAV mobility and its impact on inference latency.


While the UAV swarms are moving during the mission, studying the UAV's transmission reliability is essential in terms of mitigating the latency. The transmission reliability of UAVs has been studied in \cite{zeng2018wireless}. The authors designed a system model of UAVs that follow a master UAV, which provides instructions to the other UAVs. They set a latency threshold, and then explore the reliability of transmission based on the threshold. Moreover, the master UAV controls all the missions, and in case of its disruption, the swarm will collapse, and the mission will be terminated. 
In the studies \cite{chen2021urllc}, UAV resources have not been investigated affecting the results. In this paper, we calculate the transmission power of UAVs at the needed reliability level, and then the optimal positions of UAVs are estimated. Finally, the latency will be minimized through optimal allocation of the layers among the available resource-limited UAVs.

In this paper, we study the deployment of Convolutional Neural Networks (CNNs) within a multi-UAV surveillance system. Our strategy involves distributing CNN layers across multiple UAVs, which minimizes the latency while acting within the constraints of the participating devices as well as the distance between them. 
To the best of our knowledge, we believe that our work is the first to use UAVs for surveillance-based applications assisted by CNN distribution while considering the mobility of devices to improve reliability and reduce latency. Our paper's contributions are stated as follows:
\begin{itemize}
    \item We develop a distributed inference system consisting of a swarm of UAVs working together in  a mission to capture images and process them using a CNN-based classifier. Our system respects the resource constraints of the UAVs in swarm (i.e., computation, memory, energy), and the UAVs' positions. 
    \item We formulate the system model as an optimization problem that aims to improve the reliability and reduce the latency while distributing the inference tasks into the connected UAVs. We find the optimal transmit powers of UAVs, the optimal positions of UAVs, and the optimal layer allocations that minimize the classification latency and improve transmission reliability.
    \item We evaluate our system model through extensive simulations to prove its performance by considering different CNN networks, different device configurations, and different numbers of participants. We also compare our LLHR model with two competitive works, heuristic and random selection baseline, and show that our model outperforms the two competing models.
\end{itemize} 
This paper is organized as follows: section \ref{info_system_model} shows the system model. Section \ref{problem_formulation} presents the problem formulation, while section \ref{Performance_evaluation} shows the performance evaluation and the simulation results. Finally, section \ref{conclusion} presents the conclusions and future work.
\section{system model}
\label{info_system_model}
\begin{figure}[!ht]
    \centering
    \includegraphics[width=0.45\textwidth]{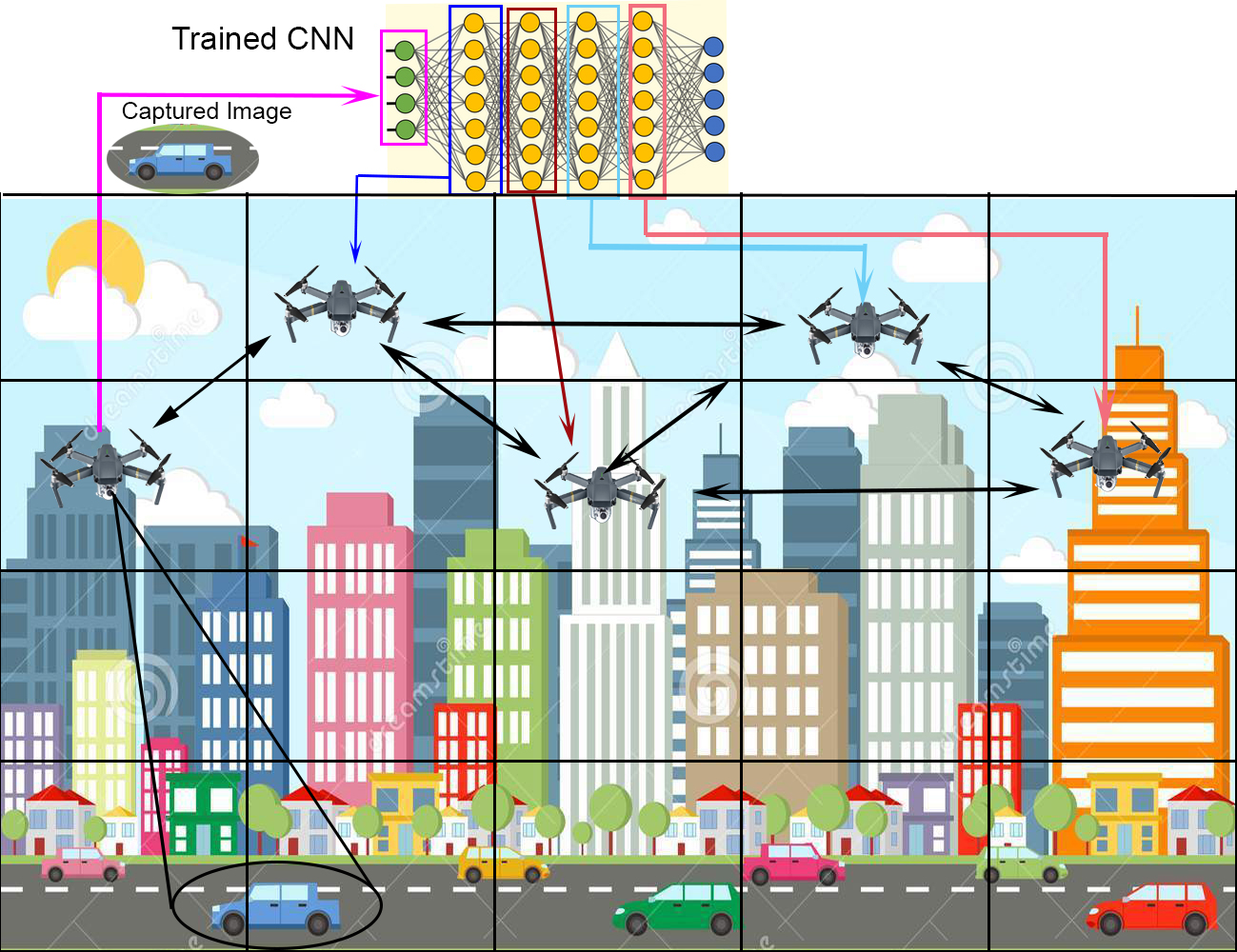}
    \caption{System Model.}
    \label{System_Model}
\end{figure}

Our system model is shown in Figure \ref{System_Model}. In this approach, we assume that a group of UAVs, denoted by $U$, is capable to form a wireless $ ad\;hoc$ network. These UAVs cover an area by capturing images from the ground and sending them to the CNN for image classification. Each UAV captures multiple images in real time and executes part of the requests. To achieve the mission within the UAV swarm, the UAV swarm collaborates to compute the classification prediction of the captured images. The UAVs transmission network is assumed to assist in low latency and high-reliability data transmission. Our approach is adequate for multi purposes like surveillance (e.g., expected forest fire areas and disaster scenes that humans cannot come close to) and tracking an object on the ground. 
The monitored area is divided into $v \times q$ cells, and each UAV navigates to cover the area. Considering the energy limitation of each UAV, the maximum power that the UAV can transmit is denoted by $P_{max}$. We assume that all UAVs exist at the same altitude; hence the location is represented as the coordinates $(x_i, y_i)$.

The captured requests are divided into subtasks, and each UAV is dedicated to executing one subtask. Supposing any UAV does not have enough resources to accomplish its assigned subtask, then it will delegate this subtask to another UAV to execute it until the whole request is completed. The monitored area might include various applications of surveillance systems 
which means multiple CNN models could be used based on the purpose of the mission.
\subsection{CNN Network Configuration}
The model comprises $U$ connected UAVs that capture requests from the ground and feed them into the CNN network. This system model seeks to find the set of UAVs that achieves the lowest latency and highest reliability, i.e., minimizing the latency of finding the final classification prediction and minimizing the total power at which the UAVs can reliably share the data. Considering the capacity constraints of UAVs, each UAV $i$ has a maximum memory $\bar{m_i}$ and maximum computational capacity $\bar{c_i}$. In this model, at each time frame $T$, each UAV $i \in \{ 1, \dotsc ,U \}$ is able to generate $RQ_i$ requests such that $\sum_i{RQ_i} = RQ$ where $RQ$ is the total number of requests at time frame $t$.

According to this scenario, the UAVs capture images, and a trained CNN network is used for classifying the input inference requests. Each UAV has a copy of the trained CNN network and can execute one subtask. The distributed system is compatible with the typical CNN networks designed for a pipeline of successive layers (e.g., convolutional and ReLU) \cite{he2016deep}. Let us denote $L$ as layers of the CNN model, which is the number of subtasks divided and distributed to the connected UAVs to classify the input captured image. Particularly, $j$ is the index of the layers where $j \in \{1, \dotsc , L \} $. Each CNN layer has a memory constraint $m_j$ and computational capacity constraints $c_j$. The memory usage $m_j$ (in bytes) represents the number of weights that the layer $j$ can store multiplied by the size of the data type dedicated to characterizing the parameters. The computational capacity $c_j$ represents the number of multiplications processed to execute the assigned layer $j$. Therefore, the computational load requirement for the layer $j$ is calculated as follows:
    \begin{equation}
         c_{j} = n_{j-1}\;.\;s_{j}^2\;.\;n_j \;.\;z_{j}^2
        \label{1st_eqn}
    \end{equation}
where $n_{j-1}$ is the number of the input channels at layer $j$, which represents the total feature maps that come from the layer $j-1$, $s_j^2$ and $z_j^2$ are the spatial filter size of the $j$-th layer and the spatial size of the output map, respectively. The computational load calculations of the fully-connected layer $j$ are the number of neurons $n_j$ of layer $j$ multiplied by the number of neurons $n_{j-1}$ of the layer $j-1$, and it is represented as follows:
    \begin{equation}
         c_{j} = n_{j-1}\;.\;n_j
        \label{2nd_eqn}
    \end{equation}
The memory requirements of layers are calculated as the number of weights $W_j$ of the current layer $j$ by the number of bits $b$ dedicated to storing the weight, and it is represented as:
    \begin{equation}
         m_{j}^{N_c} = W_j\;.\;b
        \label{3rd_eqn}
    \end{equation}
\subsection{Minimum Achievable Data Rate for Reliable Transmission}
To achieve the requirements of reliable transmission of intermediate data between UAVs, the transmission power of each UAV is minimized based on the UAV's distances. This transmission power should be less than a specified $P_{max}$ and higher than a threshold designed based on the UAV's positions. Due to the direct link between UAVs, we assume that the transmission of data between UAVs meets the Line of Sight (LoS) model and ignore the Non LoS (NLoS) in this model. The power model in each UAV is represented as $P_{i,k} = h_{i,k}. P_i$, where $h_{i,k}$ is defined as the channel gain from UAV $i$ to UAV $k$, and $P_i$ is the transmission power from UAV $i$. The channel gain is calculated as follows:
 \begin{equation}
    \small
         h_{i,k} = 
          \frac {h_0}{d(i,k)^2} \; . \; \forall i, k \in U
        \label{channel_gain}
    \end{equation}
where $h_0$ is defined as the median of the mean path gain at reference distance $d_0 = 1$ m, $d(i,k)$ represents the distance between the the UAV $i$ and UAV $k$ and it is calculated using the Euclidean distance as $\sqrt{(x_i-x_k)^2+(y_i-y_k)^2}$.
When UAV $i$ sends intermediate data to UAV $k$, the lower bound (lb) achievable data rate can be calculated as follows:
    \begin{equation}
         \rho_{i,k}(lb) = B_{i,k}\: log_{2}(1+\frac{P_{i,k}}{\sigma^2})
        \label{data_rate}
    \end{equation}
where $\sigma^2$ is the thermal noise power, and $B_{i,k}$ is the transmission bandwidth between UAV $i$ to UAV $k$.
\section{problem formulation}
\label{problem_formulation}
Considering the resource limitations of UAVs, we aim to use the minimal amount of the UAVs' power that will ensure high reliability and minimum end-to-end latency. Our system model is designed to deal with real-time requests, in which it aims to find the optimal transmission power that guarantees higher reliability and lower latency of the system, find the optimal positions of the connected UAVs, and find the optimal placement of the CNN layers in the available UAVs. We formulate this scenario into three sub-problems; the first one aims to find the optimal transmit power of the connected UAVs based on the positions of the distributed UAVs in the grid. The output of the first sub-problem is the input to the second sub-problem that aims to find the optimal positions of UAVs based on the maximum coverage area that the UAV can cover. The outputs of the first and second sub-problems are the input to the third sub-problem, which aims to find the optimal positions of layers. The following sections explain each sub-problem.
\subsection{Transmit Power Optimization}
The first sub-problem aims to find the optimal transmit power of the connected UAVs in the swarm at which the reliability of transmission is assured to be satisfied. At this step, the optimal transmit power should be less than the maximum power $P_{max}$ and greater than a threshold that is based on the lower bound of achievable data rates. The driven threshold is set to assure that the data transmission is reliable among the connected UAVs.
The optimization problem of the first sub-problem is formulated as follows:
\begin{equation}
    \begin{aligned}
    \textbf{P1: } \min_{p_i} \sum_{i=1}^U p_i
    \end{aligned}
    \label{P1}
\end{equation}
s.t.
\begin{equation}\tag{6a}
    \begin{aligned}
        P_i \geq P_i^{th} , \; \forall i \in U 
    \end{aligned}
\end{equation}
\begin{equation}\tag{6b}
    \begin{aligned}
        0 \le p_i \le p_{max}
    \end{aligned}
\end{equation}
where $P_i^{th}$ is the threshold transmit power for the UAV to transmit a data packet of size $K_j$ from one UAV to another under the reliability requirements of data transmission rate, which is calculated by using the equation of $R_{lb} . \tau = K_j$ into equation \ref{data_rate}, and we get:
\begin{equation}
    \begin{aligned}
        P_i^{th} = \frac{ \sigma^2}{h_{i,k}} [\exp{(\frac{K_j.\ln(2)}{B_{i,k} . \tau})}-1]
    \end{aligned}
\end{equation}
Note that problem \textbf{P1} in equation (\ref{P1}) is a Mixed Integer Linear program (MILP) which is a convex problem in terms of the transmit power. In order to solve problem \textbf{P1} and find the global optimum, a combination of traditional convex optimization algorithms with simple algorithms like exhaustive search can be used to obtain the optimal solution when the optimization limits are set properly.

\subsection{UAVs Positions Optimization}

The second sub-problem is to find the optimal positions of the connected UAVs in the grid. 
Let us denote the decision variable of coordinators as $S_i$ which refers to the optimal UAVs positions in the grid where $S_i = \{ \{ x_i, y_i\}, \forall i \in U\}$. The optimization problem to find the UAVs positions with minimum transmit power is formulated as follows:
\begin{equation}
    \begin{aligned}
    \textbf{P2: } \min_{P_i, S_i} \sum_{i=1}^U P_i
    \end{aligned}
\end{equation}
s.t.
\begin{equation}\tag{8a}
\small
    \begin{aligned}
        P_i \geq P_i^{th} , \; \forall i \in U 
    \end{aligned}
    \label{power_threshold}
\end{equation}
\begin{equation}\tag{8b}
\small
    \begin{aligned}
        0 \le p_i \le p_{max}
    \end{aligned}
    \label{max_power_threshold}
\end{equation}
\begin{equation}\tag{8c}
\small
    \begin{aligned}
    x_i^2 + y_i^2 \le R^2
    \end{aligned}
    \label{circle_const}
\end{equation}
\begin{equation}\tag{8d}
    \begin{aligned}
    \small
    d_{i,k} \geq 2R, \; \forall i,k, \in U
    \end{aligned}
    \label{min_distance}
\end{equation}
The constraints in equation (\ref{max_power_threshold}) is set to assure that the UAVs power is under $P_{max}$ at which the reliable transmission is assured.
The constraint in equation (\ref{circle_const}) is set to assure that the positions of the UAVs are within a circle; however, the constraint in equation (\ref{min_distance}) is set to indicate the minimum distance between UAVs in order to avoid collisions between UAVs and make sure that UAVs are flying in a safe enough distance from each other.
In order to guarantee reliable transmission and achieve low latency between UAVs, the threshold designed for each UAV should not exceed the maximum power allowable to the UAVs. Hence, the optimal solution of the problem is obtained when the equality is satisfied in equation (\ref{power_threshold}), and the optimization problem \textbf{P2} can be obtained by solving the following optimization problem:
\begin{equation}
    \begin{aligned}
    \min_{S_i} \sum_{i=1}^U \frac{\sigma^2}{h_0}[\exp{(\frac{K_j.\ln(2)}{B_i . \tau})}-1]d_{i,k}^2
    \end{aligned}
    \label{P2}
\end{equation}
s.t.
\begin{equation}\tag{9a}
    \begin{aligned}
    \small
        \frac{\sigma^2}{h_0}[\exp{(\frac{K_j.\ln(2)}{B_i . \tau})}-1]d_{i,k}^2 \le p_{max}, \; \forall i,k \in U
    \end{aligned}
    \label{threshold_Max_power}
\end{equation}
\begin{equation}\tag{9b}
    \begin{aligned}
    (\ref{circle_const}) - (\ref{min_distance})
    \end{aligned}
\end{equation}
The problem in equation (\ref{P2}) is a quadratically constrained quadratic program (QCQP) problem with respect to $d_{i,k}$, and it guarantees the threshold drawn in \textbf{P1} in equation (\ref{P1}) not to exceed the maximum power of each UAV.

\subsection{Layers Allocations Optimization}

The third sub-problem aims to distribute the sub-tasks of the captured requests into the available UAVs and choose the UAV that achieves the best latency. If any UAV does not have enough capacity, it delegates the sub-task to the next UAV to execute this sub-task. The collaboration of UAVs allows the system to accomplish the mission within the UAVs swarm without the need for transmitting it to remote servers. In this sub-task, we minimize the latency of calculating the final classification prediction, where at this point, each UAV is set in its optimal position and has its optimal power that guarantees the highest reliability. Let us define a decision variable $\delta_{r,i,j}$, where:
    \begin{equation}
    \begin{aligned}
    \small
        \delta_{r,i,j} 
        = \left\{ 
        \begin{array}{l}
            1 \;\text{if UAV $i$ executes the layer $j$ of request $r$}\\
            0 \; \text{otherwise}\\
        \end{array}
        \right.
        \end{aligned}
        \label{Deciosion_Variable}
    \end{equation}
The objective function minimizes the latency of finding the final classification prediction of the captured image is shown in equation (\ref{P3}). We emphasize that we adopted the method of per-layer distribution as it is suitable for systems with a limited number of UAVs. Finally, to support the dynamics of the system over time, the optimization is executed periodically.

Our problem is an Integer Linear programming (ILP) optimization and it is formulated as follows:
\begin{equation}
    \begin{aligned}
    \small
    \textbf{P3: }
        \min_{\delta_{r,i,j}}
        \sum_{r=1}^{RQ}
        \sum_{i=1}^{U}
        \sum_{k=1, k\neq i}^{U}
        \sum_{j=1}^{L-1}
        \delta_{r,i,j}\;.\;
        \delta_{r,k,j+1}\\ \;.\;\frac{K_{j}}
        {\rho_{i,k}} 
        \;+
        \sum_{i=1}^{U} t_i ^{(p)} \;+
        t_s
        \end{aligned}
        \label{P3}
    \end{equation}
    s.t.
    \begin{equation}\tag{11a}
    \small
        \sum_{r=1}^{RQ}
        \sum_{j=1}^{L} 
        \delta_{r,i,j} \;. \;m_{j} 
        \le 
        \bar{m_i}, \quad
        \forall i 
        \in{U},
        \label{mem_const}
    \end{equation}
    \begin{equation}\tag{11b}
    \small
        \sum_{r=1}^{RQ}
        \sum_{j=1}^{L} 
        \delta_{r,i,j} \;. \; c_{j} 
        \le 
        \bar{c_i}, 
        \forall i 
        \in{U},
        \label{comp_const}
    \end{equation}
    \begin{equation}\tag{11c}
    \small
        \sum_{i=1}^{U} 
        \delta_{r,i,j}  
        = \left\{ 
        \begin{array}{l}
            1 \;\text{\;if $j$ $ \le $ $L$}\\
            0 \; \text{\;otherwise}\\
        \end{array},
        \right. \quad
         \forall r 
        \in {RQ}, 
        \forall j 
        \in {L},;
        \label{one_cell_const}
    \end{equation}
          \begin{equation} \tag{11d}
          \small
        \delta_{r,i,j} \in \{0,1\} \; ,
         \forall j 
        \in {L},
        \forall i 
        \in {U},
        \forall r 
        \in{RQ}
        \label{binary_cons}
    \end{equation} 
and where
\begin{equation}
    \small
        t_s =
        \sum_{r=1}^{RQ}
        \sum_{k=1,  k\neq s}^{U}
       \bar\delta_{r,s,1}.\delta_{r,k,1}. \frac{K_s}{\rho_{i,k}}.
         \label{source_eqn}
    \end{equation}
    \begin{equation}
    \small
        t_i^{(p)} =
        \sum_{r=1}^{RQ}
        \sum_{j=1}^{L}
        \delta_{r,i,j} \;.
         \frac{c_{j}}{e_i}
         \label{layer_processing_eqn}
    \end{equation}
Equation (\ref{P3}) calculates the total latency of making the final classification of the captured images, in which it consists of three main parts:
\begin{itemize}
\item 
The latency $t_s$ in equation (\ref{source_eqn}) is the time required by the UAV that captures the request from the ground to transmit its collected data to the UAV computing the first layer of this request. $K_s$ is the size of the data collected by the source UAV $s$ from the ground. In equation (\ref{source_eqn}), $\bar{\delta}_{r,i,1}$ represents the complement of the decision variable $\delta_{r,i,1}$. 

\item $t_i^{(p)}$ is the total time required to compute all tasks assigned to device $i$, which is presented in equation (\ref{layer_processing_eqn}). This time is defined as the ratio between the computational requirement $c_{j}$ of the layer to the number of multiplications $e_i$ that UAV $i$ can process in a second.

\item The time required to transmit the intermediate representation of the input image from the UAV device $i$ to the UAV device $k$ that are assumed to be connected in the swarm and can be represented in equation (\ref{Tx_time}) as shown below:
    \begin{equation}
    \small
        \frac{K_{j}}{\rho_{i,k}}
        \label{Tx_time}
    \end{equation}
where $K_{j}$ is the intermediate data generated from layer $j$, $\rho_{i,k}$ represents the transmission data rate between the UAV device $i$ and the UAV device $k$. It represents the distance and the quality of the transmission connection between UAVs. Since UAVs are moving, the value of $\rho_{i,k}$ is changing according to the distance between the UAVs over the time interval in the optimization. The data rate is calculated according to equation (\ref{data_rate}).
\end{itemize}

The constraints in equations (\ref{mem_const}) and (\ref{comp_const}) are added to set a threshold for the maximum memory usage and the maximum computational capacity of the UAVs in the swarm, respectively. These thresholds are set to make the UAVs work within the allowable limit that respects their resources and to make the latency calculation compatible with the real models. Moreover, the constraint in equation (\ref{one_cell_const}) is added to ensure that each layer is computed by one UAV. This condition is added to avoid collision between different UAVs. Constraints in equation \ref{binary_cons} refer to the binary indication of the decision variables and they accept only the value of 0 or 1. 
\section{SIMULATION RESULTS AND ANALYSIS}
\label{Performance_evaluation}
Our approach uses a surveillance-based scenario for evaluation. In the experiments, we used two different CNNs, a small CNN, namely LeNet (2 convolutional layers and 3 fully connected layers), which is trained with 32×32×3 RGB-sized image, and a medium-sized CNN, namely AlexNet (5 convolutional layers and 3 fully connected layers), which is trained with a 227×227×3 RGB-sized image. Each UAV has a camera to capture an image from the monitored area and launch the inference request. We adopted an area of 480m × 480m to be covered by the UAV swarm, in which we have 144 equal size cells with 40m width and 40m length with radius $R$ of 20m. Furthermore, we used three different types of UAVs, which have the resource capabilities of the family of Raspberry Pi B+ \cite{daryanavard2018implementing}. Particularly, all types are characterized by a 1.4 GHz 64-bit quad-core processor and 1 GB RAM, while they can perform a different number of multiplications per second $e_i$ \cite{disabato2019distributed} which are 560, 512, and 256. The thermal noise $\sigma^2$ is set to -170 dBm, the mean path gain $h_0$ is $10^{-5}$, $P_{max}$ is 120 mW, and transmission duration of data packets $\tau$ is $10^{-4}$.

\begin{figure}[!t]
    \centering
    \includegraphics[width=0.45\textwidth]{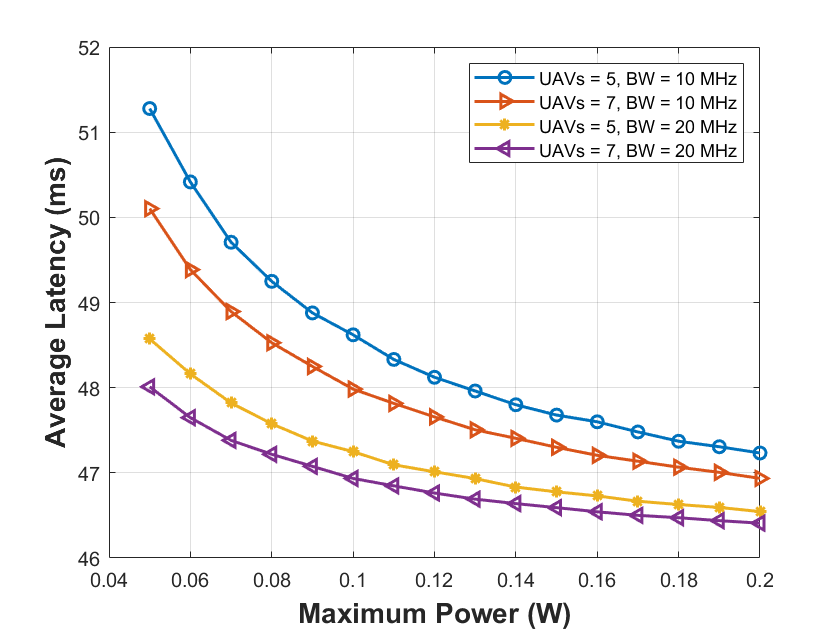}
    \caption{Average Latency for different UAVs and Bandwidths and UAVs.}
    \label{fig:diff_Max_p}
\end{figure}

Figure \ref{fig:diff_Max_p} shows the results of latency in case of changing the maximum power of each UAV under a different number of UAVs and different allocated bandwidth to the UAVs to reliably transmit intermediate data. As the maximum power of each UAV increases, the total average latency decreases since data can be sent reliability. Furthermore, as the number of UAVs increases, the total latency decreases simultaneously since the system has more flexibility in the distribution. In the same experiment, we studied the case of different bandwidth allocations between the connected UAVs, namely 10 MHz and 20 MHz, and the results showed that as we increase the bandwidth allocation, the latency decreases, and this is because the ability to transmit data reliably increases.

\begin{figure}[!t]
\centering
	\mbox{
	    \hspace{-7mm} \subfigure[\label{im8}]{\includegraphics[scale=0.3]{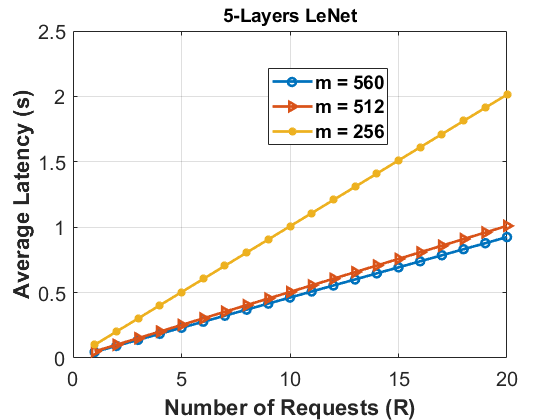}
	   }
	     \hspace{-7mm}
	     \subfigure[\label{imnomb}]{\includegraphics[scale=0.3]{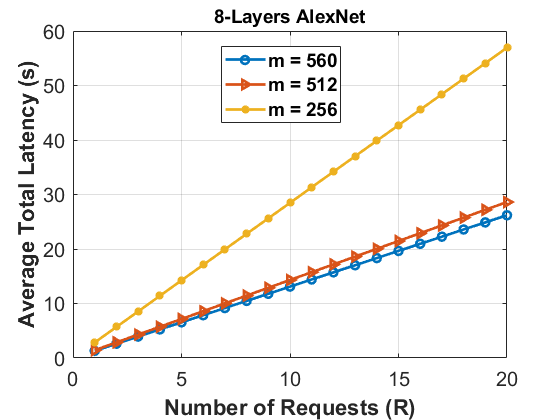}}
} 
	\caption{Average Latency for 5-layers LeNet and 8-layers AlexNet}
	\label{fig:diff_devices}
\end{figure}

Figure \ref{fig:diff_devices} shows the average latency in the case of using two different CNN networks, namely 5-layers LeNet in Figure \ref{fig:diff_devices} part (a) and 8-Layers AlexNet in Figure \ref{fig:diff_devices} part (b), and in the case of 3 different versions of Raspberry Pi 3B+ devices. As the memory capacity increases, the latency decreases in both versions of the CNN network. Moreover, as the number of requests increases, the latency increase at the same time.

\begin{figure}[!t]
\centering
	\mbox{
	    \hspace{-7mm} \subfigure[\label{im9}]{\includegraphics[scale=0.2]{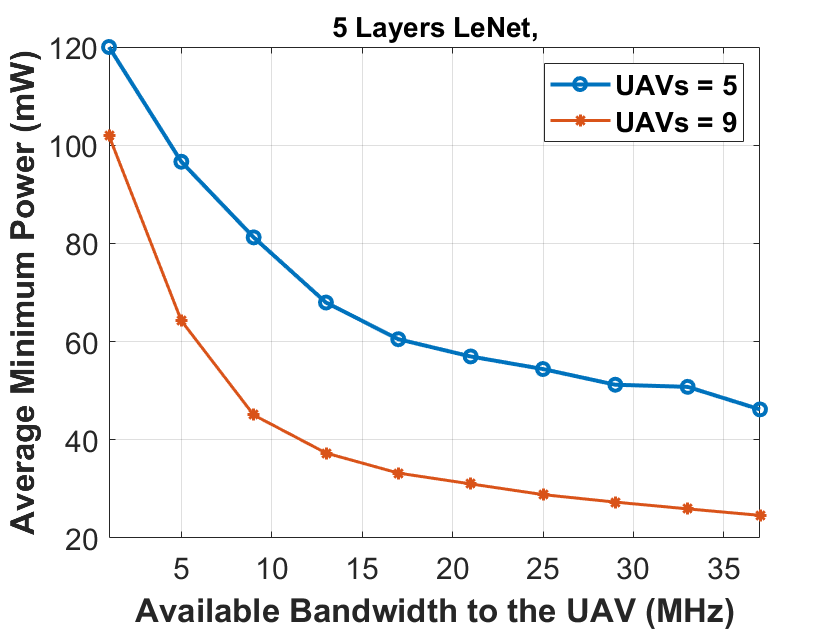}
	   }
	     \hspace{-7mm}
	     \subfigure[\label{im10}]{\includegraphics[scale=0.2]{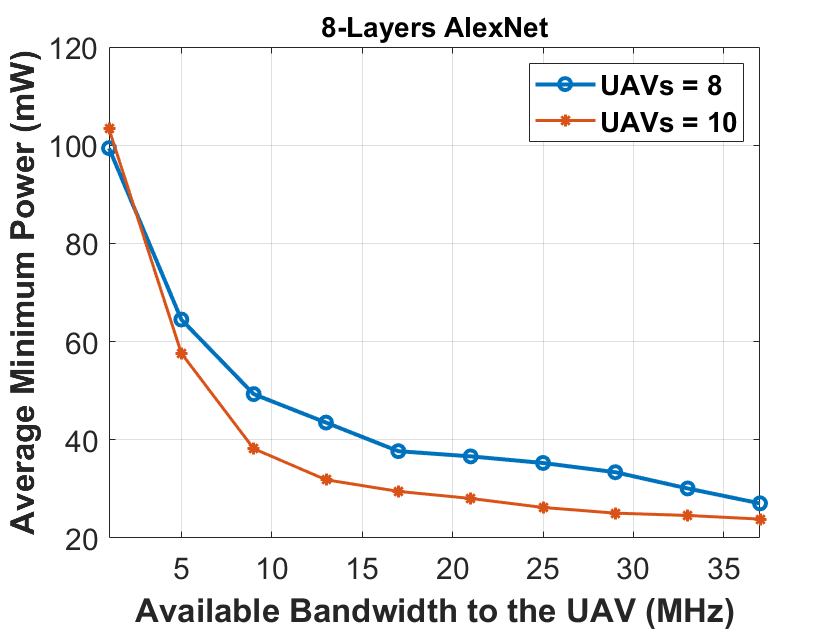}}
} 
	\caption{Average Minimum Power for 5-layers LeNet and 8-layers AlexNet in Different UAVs}
	\label{fig:min_power}
\end{figure}

Figure \ref{fig:min_power} shows the results of average minimum power to reliably transmit intermediate data between UAVs in the case of using different bandwidth allocations, different numbers of UAVs, and different CNN networks. As the bandwidth allocations increase, the average minimum power decreases in LeNet (Figure \ref{fig:min_power} part (a)) and AlexNet (Figure \ref{fig:min_power} part (b)). Furthermore, as the number of UAVs increases, the average minimum power decreases simultaneously.

\begin{figure}[!t]
    \centering
    \includegraphics[width=0.45\textwidth]{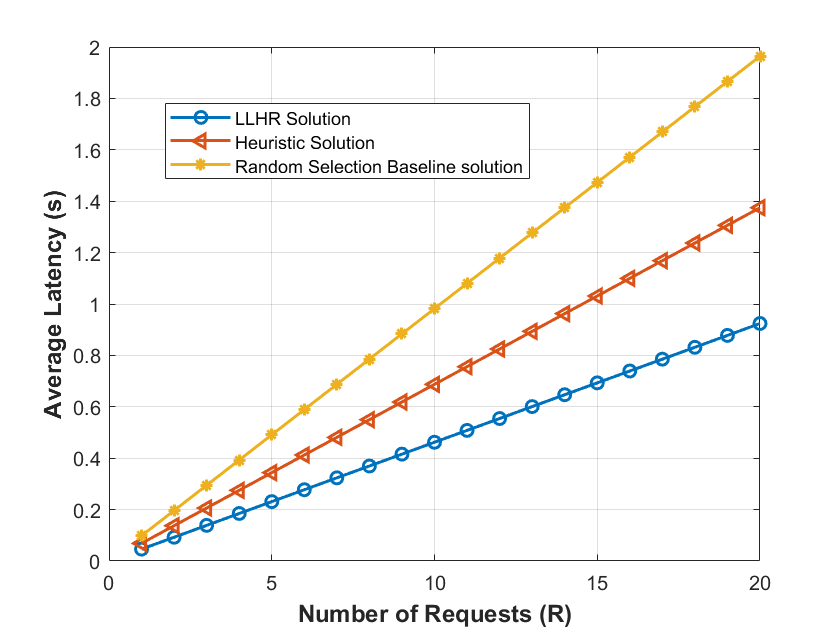}
    \caption{Average Latency for Different Requests in Different Models.}
    \label{fig:Modelscomparisons}
\end{figure}

Figure \ref{fig:Modelscomparisons} compares our LLHR approach with two other approaches, one is a heuristic solution, and the other is a random selection baseline approach as we vary the number of requests. It is clear that our LLHR approach is outperforming the other approaches. In the heuristic solution, the system model configuration is the same as the LLHR model, except that the UAVs have a static path to follow that is defined in the input configuration. In contrast, our LLHR approach has the flexibility to choose the UAV paths that minimize the UAV's power for reliable transmission between them and also gives the best latency calculations to accomplish the overall task execution. In the random selection baseline approach, the UAVs randomly move in the covered area, which produces the worst latency compared to the LLHR approach and heuristic approach as well.
\section{conclusion}
\label{conclusion}
In this paper, we developed an innovative distributed inference system in a UAV swarm while planning the optimal trajectory path for low latency and high-reliability transmission between UAVs. The system model is designed to overcome the limitations of classifying the captured image in remote servers by breaking up the request into tasks and executing these tasks locally in the UAV swarm by exploiting the limited capacities of the connected UAVs. Simultaneously, the UAVs plan their path to cover the monitored region. We formulated the distributed inference model as an optimization problem that seeks to minimize the latency of making the final classification decision and minimize the UAV's transmit power, while guaranteeing the reliability and finding the best UAV path to mitigate interference. Our simulation results demonstrated the effectiveness of our proposed solution in minimizing the latency and total transmission power. In our future work, we plan to study the effect of interference on transmission latency when controlling channel communication under different modes like a line of sight and a non-line of sight.
\section*{Acknowledgment}
This work was made possible by NPRP grant \# NPRP13S-0205-200265 from the Qatar National Research Fund (a member of Qatar Foundation). The findings achieved herein are solely the responsibility of the authors.
\bibliographystyle{IEEEtran}
\bibliography{bibliography.bib}

\begin{thebibliography}{10}
\providecommand{\url}[1]{#1}
\csname url@samestyle\endcsname
\providecommand{\newblock}{\relax}
\providecommand{\bibinfo}[2]{#2}
\providecommand{\BIBentrySTDinterwordspacing}{\spaceskip=0pt\relax}
\providecommand{\BIBentryALTinterwordstretchfactor}{4}
\providecommand{\BIBentryALTinterwordspacing}{\spaceskip=\fontdimen2\font plus
\BIBentryALTinterwordstretchfactor\fontdimen3\font minus
  \fontdimen4\font\relax}
\providecommand{\BIBforeignlanguage}[2]{{%
\expandafter\ifx\csname l@#1\endcsname\relax
\typeout{** WARNING: IEEEtran.bst: No hyphenation pattern has been}%
\typeout{** loaded for the language `#1'. Using the pattern for}%
\typeout{** the default language instead.}%
\else
\language=\csname l@#1\endcsname
\fi
#2}}
\providecommand{\BIBdecl}{\relax}
\BIBdecl

\bibitem{bejiga2017convolutional}
M.~B. Bejiga, A.~Zeggada, A.~Nouffidj, and F.~Melgani, ``A convolutional neural
  network approach for assisting avalanche search and rescue operations with
  uav imagery,'' \emph{Remote Sensing}, vol.~9, no.~2, p. 100, 2017.

\bibitem{wu20205g}
Q.~Wu, J.~Xu, Y.~Zeng, D.~W.~K. Ng, N.~Al-Dhahir, R.~Schober, and A.~L.
  Swindlehurst, ``5g-and-beyond networks with uavs: From communications to
  sensing and intelligence,'' \emph{arXiv preprint arXiv:2010.09317}, 2020.

\bibitem{al2018survey}
A.~Al-Kaff, D.~Martin, F.~Garcia, A.~de~la Escalera, and J.~M. Armingol,
  ``Survey of computer vision algorithms and applications for unmanned aerial
  vehicles,'' \emph{Expert Systems with Applications}, vol.~92, pp. 447--463,
  2018.

\bibitem{padro2019comparison}
J.-C. Padr{\'o}, F.-J. Mu{\~n}oz, J.~Planas, and X.~Pons, ``Comparison of four
  uav georeferencing methods for environmental monitoring purposes focusing on
  the combined use with airborne and satellite remote sensing platforms,''
  \emph{International journal of applied earth observation and geoinformation},
  vol.~75, pp. 130--140, 2019.

\bibitem{zeng2018wireless}
T.~Zeng, M.~Mozaffari, O.~Semiari, W.~Saad, M.~Bennis, and M.~Debbah,
  ``Wireless communications and control for swarms of cellular-connected
  uavs,'' in \emph{2018 52nd Asilomar Conference on Signals, Systems, and
  Computers}.\hskip 1em plus 0.5em minus 0.4em\relax IEEE, 2018, pp. 719--723.

\bibitem{yang2020offloading}
B.~Yang, X.~Cao, C.~Yuen, and L.~Qian, ``Offloading optimization in edge
  computing for deep learning enabled target tracking by internet-of-uavs,''
  \emph{IEEE Internet of Things Journal}, 2020.

\bibitem{9498967}
M.~Dhuheir, E.~Baccour, A.~Erbad, S.~Sabeeh, and M.~Hamdi, ``Efficient
  real-time image recognition using collaborative swarm of uavs and
  convolutional networks,'' in \emph{2021 International Wireless Communications
  and Mobile Computing (IWCMC)}, 2021, pp. 1954--1959.

\bibitem{teerapittayanon2017distributed}
S.~Teerapittayanon, B.~McDanel, and H.-T. Kung, ``Distributed deep neural
  networks over the cloud, the edge and end devices,'' in \emph{2017 IEEE 37th
  International Conference on Distributed Computing Systems (ICDCS)}.\hskip 1em
  plus 0.5em minus 0.4em\relax IEEE, 2017, pp. 328--339.

\bibitem{chen2021urllc}
K.~Chen, Y.~Wang, J.~Zhao, X.~Wang, and Z.~Fei, ``Urllc-oriented joint power
  control and resource allocation in uav-assisted networks,'' \emph{IEEE
  Internet of Things Journal}, vol.~8, no.~12, pp. 10\,103--10\,116, 2021.

\bibitem{he2016deep}
K.~He, X.~Zhang, S.~Ren, and J.~Sun, ``Deep residual learning for image
  recognition,'' in \emph{Proceedings of the IEEE conference on computer vision
  and pattern recognition}, 2016, pp. 770--778.

\bibitem{daryanavard2018implementing}
H.~Daryanavard and A.~Harifi, ``Implementing face detection system on uav using
  raspberry pi platform,'' in \emph{Electrical Engineering (ICEE), Iranian
  Conference on}.\hskip 1em plus 0.5em minus 0.4em\relax IEEE, 2018, pp.
  1720--1723.

\bibitem{disabato2019distributed}
S.~Disabato, M.~Roveri, and C.~Alippi, ``Distributed deep convolutional neural
  networks for the internet-of-things,'' \emph{arXiv preprint
  arXiv:1908.01656}, 2019.

\end{thebibliography}

\end{document}